\documentclass[prl,twocolumn,showpacs,10pt,superscriptaddress]{revtex4-1}
\usepackage[english]{babel}
\usepackage{textcomp}
\usepackage{amsmath}
\usepackage{amssymb}
\usepackage{xcolor}
\usepackage{multirow}
\usepackage{mathtools}
\usepackage{float} 
\usepackage{fontenc}
\usepackage{graphicx}
\usepackage{siunitx}
\usepackage{graphics}
\usepackage{hyperref}
\usepackage{lipsum}
\usepackage{wasysym}
\usepackage{soul}
\usepackage{combelow}
\usepackage{dsfont}
\usepackage[section]{placeins}
\makeatletter

\def\bra#1{\mathinner{\langle{#1}|}}
\def\ket#1{\mathinner{|{#1}\rangle}}

\newcommand{\sign}{\text{sign}}
\newcommand{\bea}{\begin{eqnarray}}
\newcommand{\eea}{\end{eqnarray}}

\begin{document}

\title{Efficient computation of cumulant evolution and full counting statistics: application to 
infinite temperature quantum spin chains}
\author{Angelo Valli}
\affiliation{Department of Theoretical Physics, Institute of Physics, Budapest University of Technology and Economics, M\H{u}egyetem rkp. 3., H-1111 Budapest, Hungary}
\affiliation{HUN-REN—BME Quantum Dynamics and Correlations Research Group,
Budapest University of Technology and Economics, M\H{u}egyetem rkp. 3., H-1111 Budapest, Hungary}
\author{C\u at\u alin Pa\c scu Moca}
\affiliation{Department of Physics, University of Oradea,  410087, Oradea, Romania}
\affiliation{HUN-REN—BME Quantum Dynamics and Correlations Research Group,
Budapest University of Technology and Economics, M\H{u}egyetem rkp. 3., H-1111 Budapest, Hungary}
\author{Mikl\'os Antal Werner}
\affiliation{Department of Theoretical Physics, Institute of Physics, Budapest University of Technology and Economics, M\H{u}egyetem rkp. 3., H-1111 Budapest, Hungary}
\affiliation{Strongly Correlated Systems ’Lend\"ulet’ Research Group,
HUN-REN Wigner Research Centre for Physics, P.O. Box 49, 1525 Budapest, Hungary}
\author{M\'arton Kormos}
\affiliation{Department of Theoretical Physics, Institute of Physics, Budapest University of Technology and Economics, M\H{u}egyetem rkp. 3., H-1111 Budapest, Hungary}
\affiliation{HUN-REN—BME Quantum Dynamics and Correlations Research Group,
Budapest University of Technology and Economics, M\H{u}egyetem rkp. 3., H-1111 Budapest, Hungary}
\author{\v{Z}iga Krajnik}
\affiliation{Department of Physics, New York University,  726 Broadway, New York, NY 10003, USA}
\author{Toma\v{z} Prosen}
\affiliation{Department of Physics, Faculty of Mathematics and Physics, University of Ljubljana, Jadranska 19, SI-1000 Ljubljana, Slovenia}	 
\affiliation{Institute for Mathematics, Physics, and Mechanics, Jadranska 19, SI-1000 Ljubljana, Slovenia}
\author{Gergely Zar\'and}
\affiliation{Department of Theoretical Physics, Institute of Physics, Budapest University of Technology and Economics, M\H{u}egyetem rkp. 3., H-1111 Budapest, Hungary}
\affiliation{HUN-REN—BME Quantum Dynamics and Correlations Research Group,
Budapest University of Technology and Economics, M\H{u}egyetem rkp. 3., H-1111 Budapest, Hungary}
\date{\today}


\begin{abstract}
{We propose a numerical method to efficiently compute quantum generating functions (QGF)
for a wide class of observables in one-dimensional quantum systems at high temperature. 
We obtain high-accuracy estimates for the cumulants and reconstruct 
full counting statistics from the QGF. We demonstrate its potential on spin $S=1/2$ anisotropic 
Heisenberg chain, where we can reach time scales hitherto inaccessible to state-of-the-art 
classical and quantum simulations.  Our results challenge the conjecture of the 
Kardar--Parisi--Zhang universality for isotropic integrable quantum spin chains.}
\end{abstract}

\maketitle

\emph{Introduction.---} 
A complete characterization of the time evolution of a quantum state requires
exponential resources as a function of system size, and can be carried out only for 
very small quantum systems~\cite{georgescuRMP86}. 
Therefore one typically follows the strategy of focusing on a specific observable, $\hat \Sigma$, 
often a charge or globally conserved quantity which can fluctuate in a subsytem, 
and measures its statistical properties $\Gamma \equiv \Sigma_t-\Sigma_0$ over a time period $t$. 
The corresponding distribution $P(\Gamma,t)$ is referred to as the full counting statistics (FCS) \cite{espositoRMP81}. 
To extract it, one typically performs two measurements, one at a time $t=0$, to obtain $\Sigma_0$, 
and another one at time $\Sigma_t$.  

Only few results are available for the FCS, mainly in quadratic systems \cite{levitovPZETF58,saitoPRL99,schoenhammerPRB75,doyonJPA48,yoshimuraJPA51,biaoczykSPP11,scopaJSM2022}, 
particular impurity models \cite{bagretsPRB67,belzigPRB71,gogolinPRP7,sakanoPRB83,komnikPRL107}, 
or conformal field theories \cite{bernardJSM16,grohaSPP4}, 
and only recently also in interacting 
classical~\cite{krajnikPRL128,krajnikPRR6} and
quantum~\cite{lovasPRA95,bertiniPRL131,mccullochPRL131,gopalakrishnanPRB109} 
systems. 
Experimentally, FCS has been measured in electron transport through tunnel junctions~\cite{luNat423,fujisawaAPL84,bomzePRL95,gustavssonPRL96} and quantum gases~\cite{weiSci376}. 
However, quantum computers and quantum simulators now give direct access to FCS \cite{nesterovPRA101,fanPRA109,samajdar2305.15464}, 
as recently demonstrated by the Google Quantum AI experiment~\cite{rosenbergSci384,rosenbergSci384_zenodo}, 
where the FCS of spin-transfer has been analyzed in a discrete space-time analogue of the 
spin $S=1/2$ anisotropic Heisenberg chain 
\begin{equation} \label{eq:XXZ}
 H_{\rm xxz} = J \sum_{j=1}^{L-1} \left( S^x_{j}S^{x}_{j+1} + S^y_{j}S^{y}_{j+1} + \Delta S^z_{j}S^z_{j+1} \right) 
\end{equation}
with $S^{\alpha}_j$ the $\alpha$ component of the spin $1/2$ operator, 
$J$ the spin exchange, and $\Delta$ the anisotropy parameter.  
Recently, it has been conjectured~\cite{ljubotinaNC8,ljubotinaPRL122b} that spin transport in integrable quantum spin chains 
belongs to the Kardar--Parisi--Zhang (KPZ) universality class, 
which describes space-time correlations of classical interface growth~\cite{kardarPRL56,corwinRMTA01}.  
This conundrum raise significant interest in the community~\cite{ilievskiPRL121,gopalakrishnanPRL122,denardisPRL125,denardisPRL127,scheieNatPhys17,dupontPRL127,weiSci376,yePRL129,denardisPRL131}, 
for a recent review see also Ref.~\cite{gopalakrishnanARCMP15}. 
Google's quantum simulations revealed that while spin transport is clearly anomalous at the isotropic point, $\Delta = 1$, 
and consistent with the Kardar--Parisi--Zhang (KPZ) scaling of two-point functions~\cite{ljubotinaNC8,ljubotinaPRL122b}, 
the measured skewness and kurtosis of the distribution appear to display a time evolution 
that is incompatible with those expected for the KPZ universality class. 
To validate the quantum simulations, classical quantum trajectory simulations have been performed, 
with the initial state sampled using a Monte Carlo method followed by the time evolution of each state. 
While this approach agreed with the experimental results of Refs.~\cite{rosenbergSci384,rosenbergSci384_zenodo} 
within its application range, it is unable to reach substantially longer times. 

Here we propose a Quantum Generating Function (QGF) approach, which is able to reach time scales 
so far inaccessible to classical quantum trajectory simulations in one-dimensional quantum systems, and 
evaluate cumulant to unprecedentedly long times. 
We demonstrate the method on a spin $S=1/2$ anisotropic Heisenberg model, Eq.~\eqref{eq:XXZ}, at infinite temperature and provide reliable estimates 
for the FCS and its moments beyond the limits of state-of-the-art experiments. 

\emph{The Quantum Generating Function approach.---} 
Consider a system characterized by a general density operator $\rho$
at time $t=0$. Performing a projective measurement of $\hat \Sigma$ reduces the density matrix to sectors 
associated with eigenvalues $\Sigma$ as $\rho\to \tilde \rho =\bigoplus_{\Sigma}\; \rho_\Sigma$, with 
$\rho_\Sigma = P_\Sigma\, \rho \,P_\Sigma$ the projected density operators and $P_\Sigma$ the projector to the subspace of states with eigenvalue ${\Sigma}$. 
The state described by $\tilde{\rho}$ is then evolved 
to time $t$ by the time evolution operator $U(t)$ where another projective measurement is performed
 to extract $\Sigma_t = \Sigma'$. The distribution $P(\Gamma,t)$ is given by 
 
 \begin{equation}
P(\Gamma,t) = \sum_{\Sigma,\Sigma'} \delta(\Sigma'-\Sigma- \Gamma)\; 
\mathrm {Tr} \big\{P_{\Sigma'} \,U(t) \,{\rho}_\Sigma U^\dagger(t)\big\}.
\label{eq:P(G,t)}
\end{equation}

The quantum trajectory approach samples a state from $\rho_\Sigma$, time-evolves the state and measures $\hat \Sigma$. 
The approximation of the distribution $P(\Gamma, t)$ is improved by increasing the number of samples from the initial state. 
Since the entanglement entropy of a typical initial state rapidly increases the procedure is computationally expensive. 
We circumvent this problem by directly computing the generating function of the distribution rather than the distribution itself.

\begin{figure}[b]
\includegraphics[width=1.0\linewidth, angle=0]{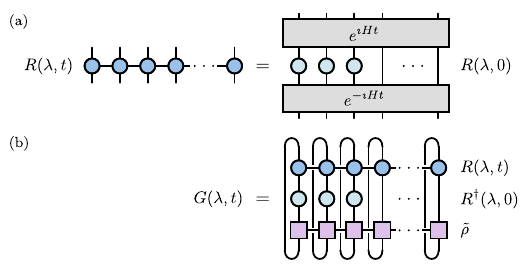}
\caption{MPO representation of (a) the unitary operator $R(\lambda,t)$ and (b) the QGF $G(\lambda,t)$. }
\label{fig:tensor_QGF}
\end{figure}

We transform Eq.~\eqref{eq:P(G,t)} into a tractable form using the 
integral representation of the Dirac $\delta$-function,  
$\delta(y) = \frac{1}{2\pi} \int_{-\infty}^{\infty} {\rm d} \lambda \;e^{i \lambda \cdot y}$, and  
introducing the QGF $G(\lambda,t)$ as 
\begin{equation} \label{eq:GRR}
  G(\lambda,t)  \equiv  \langle R(\lambda,t)R^{\dagger}(\lambda,0) \rangle_{\tilde \rho} 
\end{equation}
so that
\begin{equation} \label{eq:PFTG} 
  P(\Gamma,t) = \frac{1}{2\pi} \int_{-\infty}^{\infty} {\rm d} \lambda \; e^{-i \lambda \Gamma} G(\lambda,t)\;,
\end{equation}
where the unitary operator $R(\lambda,t) =  U(t)\, e^{i \lambda \hat\Sigma}\,U^\dagger(t) = e^{i \lambda \hat\Sigma(t)}$, 
with $U(t)$ the time-evolution operator, 
and $ \langle \dots \rangle_{\tilde \rho}  = \mathrm{Tr} \left[ \dots \tilde \rho \right]$.  
Two crucial observations are in order here. First, note that $\lim_{\lambda\to0} R(\lambda,0) = \mathop{\rm id}$, which is a trivial object of vanishing operator entanglement when represented as a Matrix Product Operator (MPO). 
At small but finite $\lambda$ the operator  $R(\lambda\approx 0,t)$ is therefore expected to have limited operator entanglement. 
The other observation is that, in most cases, the quantity of interest is a sum of commuting \emph{local} observables, 
$\hat \Sigma = \sum_{i\in \cal S} \hat O_i$, and therefore $R(\lambda,0) = \bigotimes_{i\in \cal S} e^{i\lambda \hat O_i}$, where  $\cal S$ defines a subset of sites. In this case, the MPO representation of the operator  $R(\lambda,0)$ is efficient and operator entanglement is suppressed for small $\lambda$. 
As a result, the computation of $R(\lambda\approx0,t)$ is relatively inexpensive. 
Therefore, having an MPO representation of the initial state $\tilde \rho$, 
we can evaluate the trace $\mathrm{Tr}\left[ R(\lambda,t) R^\dagger(\lambda,0) \tilde \rho \right]$ 
efficiently (see Fig.~\ref{fig:tensor_QGF}). 

The generating function gives access to the cumulants of the distribution,  
\begin{equation} \label{eq:knddn}
  \kappa_k(t) = (-\imath)^k \partial^k_\lambda \log G(\lambda,t) \big|_{\lambda=0}.
\end{equation}
Hence, besides the mean $\mu=\kappa_1$ and variance $\sigma^2=\kappa_2$, 
we can characterize the FCS by higher-order (standardized) moments 
i.e., skewness $\gamma_3=\kappa_3/\kappa_2^{3/2}$ 
and excess kurtosis $\gamma_4=\kappa_4/\kappa_2^2$ 
(note that henceforth we drop the explicit time dependence of $\kappa_k$ and $\gamma_k$ when convenient). 
In practice, we evaluate \textit{central moments}~\footnote{As we consider distributions with zero mean, moments and central moments coincide.} 
from a truncated Taylor expansion of $G(\lambda)$ 
at a finite value of the counting field $\lambda = |\lambda| e^{\imath\phi} \in \mathbb{C}$, which we extend to the complex plane,
\begin{equation}
 G(\lambda,t) = \sum_{k=0}^{\infty} \frac{1}{k!} |\lambda|^{k} e^{\imath k \phi} \mu_{k}(t). 
\end{equation}
Then $\mu_k$ is well approximated by the lowest order contribution $\propto |\lambda|^{-k}$. 
With suitable linear combinations of expansions for different phases $\phi=\{0,\pi/2,\pm\pi/4\}$ 
it is possible to cancel all expansion terms of order $m \neq k+4n$ ($n \in \mathbb{Z}$) 
and extract moments up to $k \leq 6$ with precision ${\cal O}\left( \lambda^{k+4} \right)$, 
from which the cumulants can be reconstructed using standard relations 
(see the Supplemental Material (SM)~\cite{SM} for the details).
Extracting the cumulants with a high enough precision requires a balance between 
increasing the maximum MPO bond dimension $M$, to control the truncation error, 
and decreasing $\lambda$, to control the accuracy of the Taylor expansion. 
By taking the Fourier transform \eqref{eq:PFTG} of $G(\lambda,t)$, we can also reconstruct the distribution $P(\Gamma,t)$.

\begin{figure*}[thp]
\includegraphics[width=1.0\linewidth, angle=0]{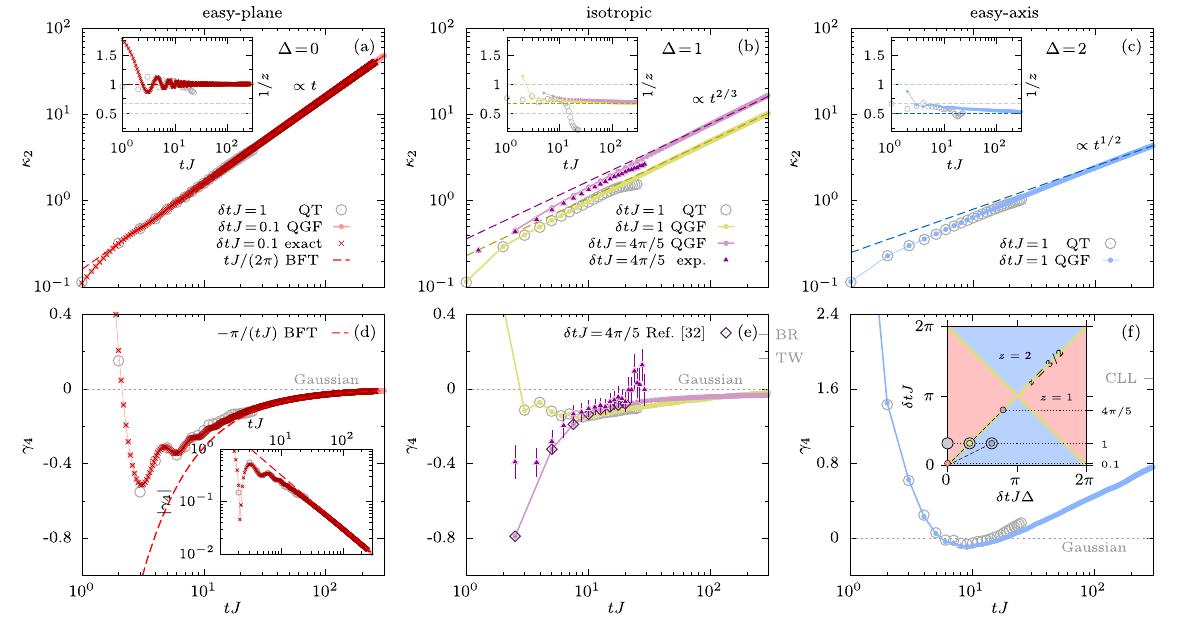}
\caption{Time evolution of the variance $\kappa_2$  
with the corresponding dynamical exponent $z$ (panels a,b,c, and insets)  
and of the excess kurtosis $\gamma_4 = \kappa_4/\kappa_2^2$ (panels d,e,f) 
in the easy-plane ($\Delta=0$), isotropic ($\Delta=1$), and easy-axis ($\Delta=2$) regimes. 
In the XX limit (a,d) QGF quasi-continuous time evolution ($\delta tJ=0.1$) is validated 
against exact numerical results and predictions of asymptotic behavior
from ballistic fluctuation theory (BFT)~\cite{myersSPP8,doyonAHP21,delvecchioJSM22}. 
At the isotropic point (b,e) the Floquet evolution ($\delta tJ=1$) is compared against 
recent experimental data~\cite{rosenbergSci384,rosenbergSci384_zenodo}. 
To compensate for different charge conventions, the $\kappa_2$ data from Ref.~\cite{rosenbergSci384_zenodo} have 
to be rescaled by a factor $r_2=32$, whereas rescaling factors cancel out in the kurtosis.  
The time step $\delta tJ=4\pi/5$ is chosen to match that of the experimental fSim gates. 
The simulation for $\gamma_4$ perfectly reproduce the exact density matrix simulations from Ref.~\cite{rosenbergSci384_zenodo}. 
Predictions for the asymptotic kurtosis $\gamma_4$ confirm the experimental conclusions 
and are inconsistent with values from the Baik--Rains (BR)~\cite{baikJSP3} and Tracy--Widom (TW)~\cite{tracyCMP159} distributions, expected in the KPZ universality class, depending on the initial condition. 
In the easy-axis regime (c,f) the statistics is strongly non-Gaussian, 
and despite the slow convergence, the kurtosis is compatible with the estimate 
form the classical Landau--Lifschitz magnet (CLL)~\cite{krajnikPRL132}. 
In all regimes, the QT simulations are performed with a time step $\delta tJ=1$. 
The schematic phases diagram (inset in panel f) identifies the 
ballistic ($z=1$ -- red), superdiffusive ($z=3/2$ -- yellow), and diffusive ($z=2$ -- cyan) 
transport regimes, whereas the symbols correspond to the QGF and QT simulations in the phase diagram. 
Parameters: $M=2^{5}$ ($\Delta=0$) and $M=2^{10}$ ($\Delta \neq 0$), $\lambda=0.03$, different $\delta t J$ (as labeled). }\label{fig:k24r_XXZ}  
\end{figure*}

\emph{Infinite temperature anisotropic Heisenberg chain.---}  
To demonstrate our method, we analyze spin transfer statistics 
in the infinite temperature anisotropic Heisenberg chain, Eq.~\eqref{eq:XXZ}.
We divide  the chain into two subsystems, ${\cal L}$ and  ${\cal R}$ of size $L/2$,
separated by an interface at bond $j=L/2$, and measure the $z$ component of the 
spin on the left, $\hat\Sigma \equiv \sum_{i\in {\cal L}} S_i^z$. 

We consider an infinite temperature state for which 
$\tilde \rho = \rho\sim \mathop{\rm id}$, and therefore $\tilde \rho$ is trivial: 
$\tilde \rho = 2^{-L} \bigotimes_{j=1}^L \mathop{\rm id}_j$. 
We can therefore recast the generating function Eq.~(\ref{eq:GRR}) as 
\begin{equation} \label{eq:Gxxz}
  G_{\rm xxz}(\lambda,t) = \frac{1}{2^L} {\rm Tr}\left[ R(\lambda,t)R^{\dagger}(\lambda,0) \right]
\end{equation}
with $R(\lambda) = \bigotimes_{j \in {\cal L}} e^{\imath \lambda S^z_{j}}$, represented as an MPO. 
 Physically, the operator $R(\lambda)$ generates a twist around the $z$ axis for spins $j \in {\cal L}$ by a small angle $\lambda$, 
 and has a simple form at $t=0$.  
The time evolved operator, 
$R(\lambda,t)=U(t)R(\lambda)U^{\dagger}(t)$ is obtained numerically by time evolving block decimation (TEBD)~\cite{vidalPRL93,vidalPRL98} using the iTensor library~\cite{itensor,itensor-r0.3}. 
The total spin,  $S^z_{\rm tot} \equiv \sum_j S^z_{j}$ commutes with $R(\lambda)$ as well as the time evolution operator $U(t)$, 
and serves as a $U(1)$ charge in TEBD simulations. 
To avoid finite size effects, we perform time evolution up to times $tJ < t_{max}J = L/2$.

\emph{Analysis of the spin $S=1/2$ model.---}
The spin $S=1/2$ XXZ model is integrable. 
Spin transport displays \textit{ballistic} and \textit{diffusive} behavior in the easy plane ($\Delta<1$) and easy axis ($\Delta>1$) 
regimes, respectively~\cite{ljubotinaPRL122a}. 
The two regimes are separated by the isotropic {Heisenberg} point ($\Delta=1$) 
where \textit{superdiffusive} transport emerges~\cite{ljubotinaNC8}. 
This phenomenology carries over to integrable Trotterized Floquet dynamics~\cite{ljubotinaPRL122a}, 
studied in Ref.~\cite{rosenbergSci384}.

Fig.~\ref{fig:k24r_XXZ} summarizes our main results; it displays the time evolution of the second cumulant $\kappa_2$ 
and the excess kurtosis, $\gamma_4 = \kappa_4/\kappa_2^2$, obtained by QGF, 
and compared with other methods as well as with the results of Refs.~\cite{rosenbergSci384,rosenbergSci384_zenodo}. 
 
\emph{(a) The free fermion limit.}
On the left-hand side (panels (a) and (d)), we present our results 
for the  XX model at $\Delta=0$, where transport is ballistic. 
In this limit, the Hamiltonian~(\ref{eq:XXZ}) can be mapped to non-interacting fermions 
which allows us to benchmark our numerical simulations against a semi-analytical exact solution. 
Using the Jordan--Wigner transformation~\cite{fradkinPRL63} and Klich's trace formula \cite{klichAP404}, 
Eq.~\eqref{eq:GRR} can be cast as the determinant of a matrix whose dimension scales linearly with system size $L$ 
and is therefore amenable to efficient exact numerical evaluation for large systems (see the SM~\cite{SM} for details).

We find perfect agreement between the QGF method and the semi-analytical results. 
This is even more striking when we compare the \textit{local dynamical exponent}, 
$z^{-1} = (d/d\log t) \log \kappa_2(t)$ (inset of panel a.).  
The second cumulant increases asymptotically in time as $\kappa_2 \propto t^{1/z}$ with $z\to1$, 
in agreement with the prediction of Ballistic Fluctuation Theory (BFT) $\kappa_2 \asymp  J t/(2\pi)$~\cite{delvecchioJSM22,SM}.  
In Fig.~\ref{fig:k24r_XXZ}, we also present the results of the usual Quantum Trajectory (QT) approach, 
which we implemented by exploiting the $U(1)$ symmetry (see the SM \cite{SM} for details). 
{Simulations are performed with a time step $\delta tJ=1$, 
favoring faster dynamics to simulate longer timescales while retaining all universal features.
The QT approach breaks down at a times $tJ \approx 20$ due to the exploding bond dimensions, 
while the QGF method provides excellent agreement even for times as large as $tJ \approx300$. 

The excess kurtosis $\gamma_4$ approaches zero as $\gamma_4\asymp -\pi/(t J)$, also in excellent agreement with BFT,
confirming that at this free fermion point Wick's theorem applies to the joint distribution. 
This is also supported by investigating the next standardized moment, $\gamma_6=\kappa_6/\kappa_2^3$ (see SM \cite{SM}). 
Our semi-analytical results show that \textit{all} even cumulant scale linearly in time $\kappa_{2n} \asymp t$, 
in agreement with the prediction of BFT (see SM \cite{SM}) 
implying that the statistics becomes asymptotically Gaussian.

\emph{(b) The isotropic point.} 
The central panels of Fig.~\ref{fig:k24r_XXZ} focus on the isotropic point, $\Delta=1$, 
where a direct comparison with experiments is also possible. 
The isotropic point is of particular interest, as we can investigate whether the scaling of higher-order cumulants 
supports the KPZ conjecture for isotropic integrable quantum spin chains~\cite{ljubotinaPRL122b}. 

The QGF method confirms superdiffusive dynamics at the isotropic point with the KPZ exponent  
$\kappa_2 \sim t^{2/3}$, $z\to3/2$, both for time steps $\delta t\, J=1$,  
and for {$\delta t\, J=4\pi /5 \approx 2.51(3)$}\footnote{In the Floquet construction, Google displays data after each layer. 
A single step therefore corresponds to a time step $\delta t\, J/2\approx 1.26$.},  
corresponding to the fSim gates used in the Google experiments (see derivation in the SM)~\cite{rosenbergSci384}. 
Remarkably, the QGF method remains stable for times far beyond the quantum simulations of Ref.~\cite{rosenbergSci384}
as well as QT simulation times; it agrees with the experiments and QT at times, $tJ \lesssim 20$,
but clearly signals the breakdown of both at times $tJ \approx 20$.


In contrast to $\kappa_2$, the kurtosis $\gamma_4$, presented in Fig.~\ref{fig:k24r_XXZ}(e) is, 
however, clearly inconsistent with KPZ scaling~\cite{rosenbergSci384}. 
Although experimental values of $\gamma_4(t)$ become unreliable at times $tJ \gtrsim 20$,
a jack-knife estimate provides an asymptotic value $\gamma_4^{\mathrm{jack-knife}}=0.05 \pm 0.02$~\cite{rosenbergSci384},   
which is incompatible with the kurtosis from Baik--Rains~\cite{baikJSP3} and Tracy--Widom~\cite{tracyCMP159} distributions, 
characterizing phenomena in the KPZ universality class~\cite{deNardisJSM2017}. 

\begin{figure}[t!]
\includegraphics[width=1.0\linewidth, angle=0]{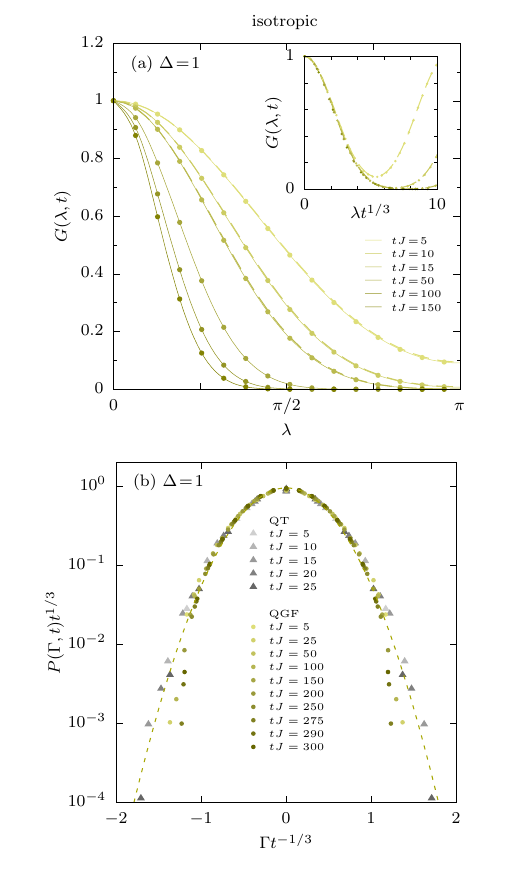}
\caption{(a) Generating function at different time snapshots obtained with QT (dashed) and QGF (solid lines), 
and their scaling collapse (insets) at the isotropic Heisenberg point ($\Delta=1$). 
(b) Scaling collapse of FCS for QT (grey gradient symbols) and QGF (color gradient symbols).  
The rescaled distribution, ${\cal P}_{1/3} (\Gamma/t^{1/3}, t) \equiv t^{1/3}\,P(\Gamma,t)$ becomes a function of 
$\Gamma / t^{1/3}$, corresponding to a dynamical exponent, $z=3/2$. 
Dashed lines represent a Gaussian distribution, for comparison.
Parameters: $M=2^{10}$, $\delta t J=1$. }
\label{fig:scaling_XXX}  
\end{figure}

Our QGF simulations confirm the overall conclusions of Ref.~\cite{rosenbergSci384}, 
but provide a refined picture over significantly longer times.
The QGF $\gamma_4$ data agree perfectly with experimental results over the first few cycles 
and with density matrix simulations of Refs.~\cite{rosenbergSci384,rosenbergSci384_zenodo}. 
However, Floquet time evolution~\cite{ljubotinaPRL122a} (performed with $\delta tJ=1$ and {$\delta tJ=4\pi/5$}) 
yield a time evolution consistent both with asymptotic Gaussian fluctuations, $\gamma_4\to 0$, or 
weakly non-Gaussian fluctuations with slightly negative kurtosis, $\gamma_4<0$.

\emph{(c) Easy-axis regime.} 
We now focus on the easy-axis regime, 
where we present simulations for $\Delta=2$ on the right panels of Fig.~\ref{fig:k24r_XXZ}.
Here $\kappa_2\propto t^{1/2}$ reveals diffusive behavior with $z=2$, but the 
numerics suggests an asymptotic kurtosis $\gamma_4>0$, corresponding to a strongly \textit{non-Gaussian} distribution. 
The kurtosis data do not show any clear sign of saturation in the investigated time window. 
This is not surprising, since even the dynamical exponent converges rather slowly towards $z=2$, see inset of Fig.~\ref{fig:k24r_XXZ}(c). 
A similarly slow convergence has been observed in a related model, 
the classical Landau--Lifshitz magnet, with asymptotic kurtosis $\gamma_4^{\mathrm{CLL}}=3(\pi/2-1)$~\cite{krajnikPRL132}, 
indicated for reference in Fig.~\ref{fig:k24r_XXZ}(f) and still compatible with our numerics. 
%

\emph{Scaling properties of the generating function and full counting statistics.---}
Our method allows to extract the generating function $G(\lambda,t)$ as a function of $\lambda$ and $t$, 
to reconstruct $P(\Gamma,t)$ by performing the Fourier transformation \eqref{eq:PFTG}. 
In Fig.~\ref{fig:scaling_XXX}, we present results for the isotropic point, $\Delta=1$; 
(the ballistic and diffusive regimes are discussed in the SM \cite{SM}). 
The curves in the Fig.~\ref{fig:scaling_XXX}(a) correspond to snapshots of $G(\lambda,t)$ at different times. 
Upon rescaling the counting field as $\lambda \to \lambda t^{1/2z}$, the data collapse to a single curve, 
as shown in the inset for the QT and QQF methods.
Analogously, Fig.~\ref{fig:scaling_XXX}(b) demonstrates the collapse of the rescaled distributions 
$t^{1/2z} P(\Gamma,t) \to {\cal P}_{1/2z}(\Gamma /t^{1/2z})$. 
In both cases, the data collapse is obtained using the appropriate dynamical exponent $z=3/2$ for $\Delta=1$. 
Although the data are not entirely decisive, at the isotropic (Heisenberg) point the tails of the distribution 
appear thinner than Gaussian, in qualitative agreement with the evolution of the kurtosis, see Figs.~\ref{fig:k24r_XXZ}(e,f). 

\emph{Conclusions and Outlook.---}
We introduced a Quantum Generating Function approach to accurately and efficiently compute 
the full counting statistics for sums of local observables and their corresponding moments.  
Our method is widely applicable to one-dimensional systems in arbitrary initial states.
We demonstrated this method on the anisotropic spin $S=1/2$ XXZ model. 
Benchmarking for $\Delta=0$ against numerically exact solutions confirmed  
the astonishing accuracy and range of validity of the QGF method. 
We are able to extract even time derivatives with high accuracy and over time scales 
more than an order of magnitude beyond those of quantum simulations in Ref.~\cite{rosenbergSci384} 
as well as quantum trajectory approaches. 
At the isotropic point, $\Delta=1$, our results are consistent with a vanishing or a weakly negative kurtosis, 
demonstrating that spin transfer statistics is inconsistent with KPZ scaling.  

\begin{acknowledgments}
This research was supported by the Ministry of Culture and Innovation and the National Research, Development and Innovation Office (NKFIH) within the Quantum Information National Laboratory of Hungary
(Grant No. 2022-2.1.1-NL-2022-00004), through NKFIH research grants Nos. K134983, K138606, SNN139581, 
and QuantERA `QuSiED' grant No. 101017733.
C.P.M. acknowledges support by the Ministry of Research, Innovation and Digitization, CNCS/CCCDI–UEFISCDI, 
under the project for funding the excellence, contract No. 29 PFE/30.12.2021. 
M.A.W. has been supported by the \'UNKP-23-2-III-BME-327 and \'UNKP-22-V-BME-330 
New National Excellence Programme of the Ministry for Culture and Innovation from the source of the 
National Research, Development and Innovation Fund, and by the Bolyai Research Scholarship of the 
Hungarian Academy of Sciences. 
\v{Z}.K. is supported by the Simons Foundation as a Junior Fellow of the Simons Society of Fellows (1141511). 
T.P. acknowledges Program P1-0402, and grants N1-0334, N1-0219, and N1-0233 of Slovenian Research and Innovation Agency (ARIS), as well as the Advanced Grant QUEST-101096208, of the European Research Council (ERC).
\end{acknowledgments}

\bibliographystyle{apsrev}
\bibliography{arXiv.bbl}

\clearpage
\newpage

\onecolumngrid

\widetext
\begin{center}
\textbf{\large Supplemental Material for:\\ Efficient computation of cumulant evolution and full counting statistics: application to 
infinite temperature quantum spin chains}
\end{center}

%
\setcounter{equation}{0}
\renewcommand{\theequation}{S\arabic{equation}}
\setcounter{figure}{0}
\renewcommand{\thefigure}{S\arabic{figure}}
\setcounter{table}{0}
\renewcommand{\thetable}{S\arabic{table}}
\setcounter{page}{1}

\tableofcontents

\newpage

\section{Full counting statistics and the generating function} \label{sec:GF}
\noindent
We present details on how to compute the generating function (QGF) of a measurable quantity of interest 
by taking advantage of the symmetries of the problem. 
The generating function gives access to the time evolution of the moments (and cumulant) of the associated probability distribution. 
The full distribution, i.e., the full counting statistics (FCS) can also be obtained by a discrete Fourier transform of the generating function.
The method that we describe is general and can be extended to models with larger number of symmetries. 
In particular, we discuss the case of the anisotropic Heisenberg chain. 

\subsection{Extract moments and cumulants of $P(\Gamma)$ from $G(\lambda,t)$ in the MPO representation}
\noindent
Let us recall that, within the QGF method that we introduced, the generating function 
of a measurable quantity of interest $\hat\Sigma$ can be evaluated as
\begin{equation} \label{eq:GRR}
  G(\lambda,t)  \equiv  \langle R(\lambda,t)R^{\dagger}(\lambda,0) \rangle_{\tilde \rho} 
\end{equation}
where the unitary operator 
\begin{equation}
 R(\lambda,t) =  U(t)\, e^{i \lambda \hat\Sigma}\,U^\dagger(t) = e^{i \lambda \hat\Sigma(t)},
\end{equation}
with $U(t)$ the time-evolution operator, and $\langle \dots \rangle_{\tilde \rho}  = \mathrm{Tr} \left[ \dots \tilde \rho \right]$ 
denoting the trace over the density matrix $\tilde\rho$ reduced to sectors spanned by the eigenvalues of $\hat\Sigma$. 

\noindent
The goal is to extract from the generating function the moments that characterize the associated FCS. 
Formally, knowing $G(\lambda,t)$ one can evaluate the (time-dependent) moments as 
\begin{equation} \label{eq:mnddn}
   \mu_{k}^{\prime}(t) = (-\imath)^k \partial^k_\lambda G(\lambda,t) \big|_{\lambda=0},
\end{equation}
whereas the cumulants re given by
\begin{equation} \label{eq:knddn}
  \kappa_k(t) = (-\imath)^k \partial^k_\lambda \log G(\lambda,t) \big|_{\lambda=0}.
\end{equation}

\subsection{Truncated Taylor expansion} 
\noindent
Numerically, rather than performing the $k$-th derivative with respect to $\lambda$, one can extract the moments 
from the Taylor expansion of $G(\lambda,t)$ at a \textit{finite} value of $\lambda$, i.e.
\begin{equation}
 G(\lambda,t) = \sum_{k=0}^{\infty} \frac{1}{k!} (\imath\lambda)^k \mu_{k}^{\prime}(t). 
\end{equation}
It is known that the generating function cannot be a finite-order polynomial of degree $k>2$. 
Hence, we evaluate moments from a truncated Taylor expansion of $G(\lambda)$ 
at a finite value of the counting field $\lambda = |\lambda| e^{\imath\phi} \in \mathbb{C}$, which we extend to the complex plane. 
The generating function is then given by  
\begin{equation}
    G(\lambda,t) = \langle R(\lambda,t)R^{\dagger}(\lambda^*,0) \rangle_{\tilde{\rho}} 
\end{equation}
and the corresponding Taylor expansion reads 
\begin{equation}
 G_{\phi}(\lambda,t) = 1 + \sum_{k=1}^{\infty} (-1)^k \frac{1}{(2k)!} |\lambda|^{2k} e^{\imath 2k \phi} \mu_{2k}^{\prime}(t),
\end{equation}
where we introduced a subscript to keep track of the phase $\phi$ of the complex counting field $\lambda$. 
Let us note at this point that, in the case of the spin transfer in the anisotropic Heisenberg chain 
$G(\lambda,t)$ is symmetric for $|\lambda| \to -|\lambda|$, and all \textit{odd} moments vanish, $\mu_{2k+1}^{\prime}=0$. 
Another consequence is that, since the mean is zero $\mu_1^{\prime}=0$, 
all moments $\mu_k^{\prime}$ and central moments $\mu_k$ coincide. 

\noindent
When limited to $\phi=0$ (i.e., purely real $\lambda$) it is possible to extract the moment $\mu_2^{\prime}$ 
from the lowest order of the expansion
\begin{equation}
  1-\Re G_{0}(\lambda,t) = \frac{1}{2!}\lambda^2 \mu_2^{\prime}(t) + {\cal O}\left(\lambda^4\right).
\end{equation}
However, the next term in the expansion $\mu_4^{\prime}$ cannot be calculated independently of $\mu_2^{\prime}$ 
from $G_{0}(\lambda,t)$. 
A suitable choice of the phase $\phi \neq 0$ allows extracting higher-order moments as well. 
For instance, for  $\phi=\pi/4$, the moments $\mu_2^{\prime}$ and $\mu_4^{\prime}$ can be extracted \textit{independently} from the imaginary and real parts of the generating function, respectively, as
\begin{equation}
 \begin{split}
  \Im G_{\frac{\pi}{4}}(\lambda,t) =& \, \phantom{1}  - \frac{1}{2!}\lambda^2 \mu_2^{\prime}(t) + {\cal O}\left(\lambda^6\right) \\
  \Re G_{\frac{\pi}{4}}(\lambda,t) =& \, 1            - \frac{1}{4!}\lambda^4 \mu_4^{\prime}(t) + {\cal O}\left(\lambda^8\right). 
 \end{split}
\end{equation}
A higher precision can be achieved by combining expansions of $G_{\phi}(\lambda,t)$ obtained for different phases $\phi$,
which allow to \textit{isolate} the moments by selectively canceling out specific terms in the Taylor expansion. 
A natural choice is to consider expansions for $\phi=0$ and $\phi=\pi$, so that following expressions yield 
both $\mu_2^{\prime}$ and $\mu_4^{\prime}$, as
\begin{equation}
 \begin{split}
  \left[\Re G_{\frac{\pi}{2}}(\lambda,t) - \Re G_0(\lambda,t)\right]/2& \,\,\phantom{-1} = \, \frac{1}{2!}\lambda^2 \mu_2^{\prime}(t) + {\cal O}\left(\lambda^6\right), \\
  \left[\Re G_{\frac{\pi}{2}}(\lambda,t) + \Re G_0(\lambda,t)\right]/2&-1 = \, \frac{1}{4!}\lambda^4 \mu_4^{\prime}(t) + {\cal O}\left(\lambda^8\right),
 \end{split}
\end{equation}
which, however, does not grant any advantage over the $\phi=\pi/4$ expansion. 

\noindent
Remarkably, this procedure can be generalized including more phases. 
The following relations for $\phi=\{0,\pi/2,\pm\pi/4\}$ are obtained by imposing the cancellation 
of all terms $m \neq k+4n$ ($n \in \mathbb{Z}$) 
and allow extracting moments up to $k \leq 6$ \textit{independently}, with precision ${\cal O}\left( \lambda^{k+4} \right)$ 
\begin{equation}
 \begin{split}
  \frac{1}{2!}\lambda^2 \mu_2^{\prime}(t) + {\cal O}\left(\lambda^{10}\right) =& -\frac{1}{4}\left[ \Im G_{\frac{\pi}{4}}(\lambda,t) - \Im G_{-\frac{\pi}{4}}(\lambda,t) - \Re G_{\frac{\pi}{2}}(\lambda,t) + \Re G_{0}(\lambda,t) \right], \\
  \frac{1}{4!}\lambda^4 \mu_4^{\prime}(t) + {\cal O}\left(\lambda^{12}\right) =& -\frac{1}{4}\left[ \Re G_{\frac{\pi}{4}}(\lambda,t) + \Re G_{-\frac{\pi}{4}}(\lambda,t) - \Re G_{\frac{\pi}{2}}(\lambda,t) - \Re G_{0}(\lambda,t) \right], \\
  \frac{1}{6!}\lambda^6 \mu_6^{\prime}(t) + {\cal O}\left(\lambda^{14}\right) =& \ \phantom{-}\frac{1}{4} \left[ \Im G_{\frac{\pi}{4}}(\lambda,t) - \Im G_{-\frac{\pi}{4}}(\lambda,t) + \Re G_{\frac{\pi}{2}}(\lambda,t) - \Re G_{0}(\lambda,t) \right].
 \end{split}
\end{equation}
This construction is particularly important when evaluating numerically higher-order cumulants, since, in the long-time limit, 
the moments grow with increasing powers of time as 
\begin{equation}
 \mu_k^{\prime} = \left\langle \Gamma^k \right\rangle \sim \left(t^{1/2z}\right)^k, 
\end{equation}
and it becomes of pivotal importance to avoid \textit{spurious} contributions from higher-order terms in each $\mu_k$ entering the expression of $\kappa_k$, 
while at the same time achieving a reasonable signal-to-noise ratio in terms of $\lambda$. 
The drawback of this method consists of the numerical overload of the multiple simulations (with different phases $\phi$) 
required for each set of parameters and, at the same time, the higher numerical precision (e.g., in terms of bond dimension) 
required to achieve a term-wise cancellation between different Taylor expansions. 

\subsection{Relation between moment and cumulants}
\noindent
In order to compare with the Google experiment, the goal is to extract the cumulants of the distribution. 
The central moments can be expressed in terms of cumulants as
\begin{equation} \label{eq:mp2k}
 \begin{split}
  \mu_1^{\prime} &= \kappa_1,\\
  \mu_{k>1}^{\prime} &= \sum_{m=1}^{k-1} \binom{k-1}{m-1} \kappa_m \mu_{k-m}^{\prime} + \kappa_k,
 \end{split}
\end{equation} 
and analogously for the central moments
\begin{equation} \label{eq:m2k}
 \begin{split}
  \mu_2 &= \kappa_2, \\
  \mu_3 &= \kappa_3, \\
  \mu_{k>2}  &= \sum_{m=2}^{k-2} \binom{k-1}{m-1} \kappa_m \mu_{k-m} + \kappa_k, 
 \end{split}
\end{equation} 
so for instance $\mu_4 = \kappa_4 + 3\kappa^2_2$ 
and $\mu_6 = \kappa_6 + 15\kappa_4\kappa_2 + 10\kappa^2_3$ + 15$\kappa^3_2$, 
where $\mu_2 = \kappa_2$ was used. 
These expressions can be inverted to extract the cumulants from the central moments. 

\newpage

\section{Quantum trajectories approach} \label{sec:QT}

\subsection{Full counting statistics from a $U(1)$ Schmidt decomposition of the time-evolution operator}
\noindent
In general, the generating function on the spin-$1/2$ chain of length $L$ for the $T=\infty$ state is given by  
\begin{equation} \label{eq:GF}
 G(\lambda)=\frac{1}{2^L}\sum_{ss'}|\!\bra{s'}U\ket{s}\!|^2 e^{\imath\lambda(\hat{\Sigma}_{s'}-\hat{\Sigma}_{s})}
\end{equation}
where $U(t)=e^{-\imath Ht}$ is the time-evolution operator connecting configurations $\ket{s}$ and $\ket{s'}$, and
$\hat{\Sigma}_s=\sum_{j \in {\cal L}} S^z_{j}(s)$ the total spin in subsystem ${\cal L}$ in configuration $\ket{s}$. 
By performing a singular value decomposition (SVD) of the time-evolution operator at the interface 
\begin{equation} \label{eq:OSVD}
 U = \sum_{\Gamma} \sum_{\alpha} \Lambda_{\Gamma\alpha} O^{\cal L}_{\Gamma\alpha} \otimes O^{\cal R}_{\bar{\Gamma}\alpha} = \sum_{\Gamma} O_{\Gamma},
\end{equation}
where $O^{{\cal L} ({\cal R})}_{\Gamma\alpha}$ acts on subsystem ${\cal L} \ ({\cal R})$, spanned by $\{\alpha\}$, 
$\Gamma$ labels the $U(1)$ sector associated with the $S_z$ component of total spin transferred across the interface 
(i.e., from ${\cal R}$ to ${\cal L}$)
and $\Lambda_{\Gamma\alpha}$ are the singular values in the corresponding total spin sector. 
Spin conservation implies $\bar{\Gamma}=-\Gamma$. 
The generating function can be recast as 
\begin{equation} \label{eq:GO}
 G(\lambda)=\sum_{\Gamma} e^{\imath\lambda\Gamma} \frac{1}{2^L}\sum_{ss'}|\!\bra{s'} O_{\Gamma} \ket{s}\!|^2. 
\end{equation}
For any given configuration pair $\{\ket{s}, \ket{s'}\}$, $\Gamma=\Sigma_{s'}-\Sigma_{s}$, and for a given $\Gamma$ the operator $O_{\Gamma}$ is non-zero only if $\Sigma_{s'}-\Sigma_{s}=\Gamma$. 
Thus, it is possible to interchange the summation over \(\Gamma\) with the summation over the configurations \((s, s')\), resulting in
\begin{equation}
 G(\lambda) = \sum_{\Gamma} e^{\imath\lambda\Gamma} \frac{1}{2^L} \sum_{ss'}|\bra{s'} O_{\Gamma} \ket{s}|^2 
            = \sum_{\Gamma} e^{\imath\lambda\Gamma} \frac{1}{2^L} {\rm Tr} \left[ O^{\dagger}_{\Gamma}\hat{O}^{\phantom{\dagger}}_{\Gamma} \right].
\end{equation}
With the unitarity condition 
\begin{equation}
 \frac{1}{2^L} {\rm Tr} \left[ {\mathop{\rm id}}^{\dagger}\mathop{\rm id} \right] = 1,
\end{equation}
where $\mathds{1}$ denotes the identity matrix. As the time-evolution operator is also normalized  
\begin{equation}
 \frac{1}{2^L} {\rm Tr} \left[ O^{\dagger}_{\Gamma}O^{\phantom{\dagger}}_{\Gamma'} \right] = \delta_{\Gamma\Gamma'} \sum_{\alpha} |\Lambda_{\Gamma\alpha}|^2.
\end{equation}
Hence, the generating function reduces to
\begin{equation} \label{eq:PbyL}
 G(\lambda) = \sum_{\Gamma} e^{\imath\lambda\Gamma} P(\Gamma) = \sum_{\Gamma} e^{\imath\lambda\Gamma} \sum_{\alpha}|\Lambda_{\Gamma\alpha}|^2.
\end{equation}
Therefore, the FCS is related to the entanglement spectrum of the subsystem~\cite{calabreseEPL129}. 
By exploiting the $U(1)$ symmetry associated with the conservation of $\hat\Sigma$, 
it is possible to evaluate full distribution $P(\Gamma)$ from the Schmidt values at the interface, 
which are readily accessible using the 
time-evolving block decimation (TEBD) approach~\cite{vidalPRL91,vidalPRL93}.

\subsection{Sampling of the initial state}
\noindent
Within the quantum trajectories (QT) approach, one performs the time evolution of an individual pure state 
\begin {equation}
 \ket{s} = \bigotimes_{j=1}^{L} \ket{\sigma_j},
\end{equation}
with $\sigma_j$ denoting magnetic quantum number. 
Each state is represented as a matrix product state (MPS). 
We employ TEBD approach 
with a brick-wall construction for which the time evolution is integrable~\cite{vanicatPRL121}. 
Decomposing the time evolution operator in even and odd layers, $U = U_{\mathrm{odd}}\cdot U_{\mathrm{even}}$, with
\begin{equation} \label{Ueo}
 U_{\mathrm{odd}}=\prod_{j=1}^{L/2} U_{2j,2j+1}, \quad U_{\mathrm{even}}=\prod_{j=1}^{L/2} U_{2j-1,2j},
\end{equation}
and $U_{j,j+1} = e^{-\imath \delta t J h_{j,j+1}}$, where $\delta t$ is the time step and 
\begin{equation}
 h_{j,j+1} = S^x_{j} S^x_{j+1} + S^y_{j} S^y_{j+1} + \Delta S^z_{j} S^z_{j+1}
\end{equation}
is the local Hamiltonian term, i.e., 
\begin{equation}
 H_{\rm xxz} = \sum_{j=1}^{L=1} h_{j,j+1}.
\end{equation} 
For each MPS, we obtain $P_j(\Gamma)$ from the singular values as in Eq.~\eqref{eq:PbyL}. 
The full counting statistics of a given \textit{ensemble} is obtained by 
Monte Carlo sampling the initial states with probability given by the corresponding density matrix
\begin{equation}
 \rho = \frac{e^{-\beta H}}{\mathrm{Tr} \left[ e^{-\beta H} \right] }.
\end{equation}
For the infinite-temperature state, the initial states are sampled with equal probability. 
For each state $\ket{s}$ we also sample 
$\ket{\bar s} = \bigotimes_{j=1}^{L} \ket{\bar\sigma_j}$ with $\bar\sigma_j=-\sigma_j$.  
The FCS distribution is then given by the Monte Carlo average over the initial states
\begin{equation}
 P(\Gamma) = \lim _{n_{MC} \to \infty} \frac{1}{n_{MC}} \sum_{n=1}^{n_{MC}} P_j(\Gamma).
\end{equation}
The distribution is found to converge rapidly with the Monte Carlo samples $n_{MC}$, 
and a few hundred samples provide statistically stable results.  
With the FCS at hand we can extract the moments as
\begin{equation}
 \mu_k^{\prime} = \sum_{\Gamma} \Gamma^k P(\Gamma) 
\end{equation}
and from those obtain the cumulants of the distribution using the relations in Eq.~(\ref{eq:mp2k}). 

\noindent
The QT approach offers several advantages, such as providing direct access to the FCS for the transferred spin across the interface. However, it has the drawback of being limited to relatively short timescales due to rapid entanglement growth. 
This leads to unfavorable scaling with the MPS bond dimension. 
While this issue can be partially alleviated by exploiting additional non-abelian $SU(2)$ symmetry~\cite{wernerPRB102},
it still imposes a significant computational constraint.

\newpage

\section{Full counting statistics of the spin-$1/2$ XX chain} \label{sec:XX}
\noindent
For $\Delta=0$ the Hamiltonian becomes that of the XX spin-$1/2$ chain of length $L \in 2 \mathbb{N}$
\begin{equation}
	H = J\sum_{j=1}^{L-1} S_j^xS_{j+1}^x + S_j^yS_{j+1}^y, \label{XX_Ham}
\end{equation}
where $J$ is the spin exchange constant, $S_j^a = \frac{1}{2} \sigma_j^a$ are spin-$1/2$ operators and $\sigma_j^a$ are Pauli matrices.
The total spin $S^z_{\rm tot} = \sum_{j=1}^L S^z_j$ commutes with the many-body propagator, $[{U}, S^z_{\rm tot}] = 0$.
The infinite-temperature spin full counting statistics of the brick-wall dynamics after $T \in \mathbb{N}$ time-steps reads 
\begin{equation}
	G(\lambda, T) =  \langle \left[{U}\right]^T e^{\imath \lambda \Sigma} \left[{U}^\dagger\right]^T 
	e^{-\imath \lambda \Sigma} \rangle_{{\beta=0}}, \label{G_def}
\end{equation}
where $\Sigma$ is the charge on half of the chain, $\Sigma = \sum_{j=L/2+1}^{L} S_j^z$.
To avoid finite-size effect we consider times up to $T \leq L/2$.

\subsection{Free fermion representation}
\noindent
We map the spin operators $S_j^\pm = S_j^x \pm \imath S_j^y$ and $ S_j^z$ 
to fermions creation and annihilation operators $a_j^\dagger, a_j$ using the Jordan-Wigner transformation
\begin{equation}
	S_j^+ = P_j a_j^\dagger/2, \quad S_j^- = P_j a_j/2, \quad S_j^z = 1/2 - n_j \qquad P_j = e^{\imath \pi \sum_{j'=1}^{j-1}n_{j'}} = \prod_{j'=1}^{j-1} (1-2 n_{j'}) = \prod_{j'=1}^{j-1} 2S_j^z, \label{JW_def}
\end{equation}
where $n_j = a_j^\dagger a_j$ is the number operator. Transformation \eqref{JW_def} maps the $\mathfrak{su}(2)$ algebra of spins $[S_j^a, S_{j'}^b] = \frac{\imath}{2} \varepsilon^{abc} S^c_j \delta_{j, j'}$ to the canonical fermion anti-commutation relations 
\begin{equation}
\{a_{j}, a_{j'} \} = \{a_{j}^\dagger, a_{j'}^\dagger\} = 0, \qquad 
\{a_j^\dagger, a_{j'}\} = \delta_{j, j'},
\end{equation}
where $\{x, y\} = xy + yx$ is the anti-commutator. We denote the one-particle fermionic states and their duals as
\begin{equation}
	|j\rangle = a_j^\dagger |0\rangle, \qquad \langle j| = \langle 0 | a_j, 
\end{equation}
where $|0 \rangle$ is the fermionic vacuum, $a_j |0 \rangle = 0$.
The Hamiltonian \eqref{XX_Ham} expressed in terms of fermions reads
\begin{equation}
	H  = \frac{J}{2} \sum_{j=1}^{L-1} a_j^\dagger a_{j+1} + a_{j+1}^\dagger a_j = J \sum_{i, j=1}^L a_i^\dagger \mathcal H_{ij} a_j, \label{XX_ham_ferm}
\end{equation}
where $\mathcal H_{ij} = J^{-1}\langle i |H|j\rangle = \frac{1}{2}\left(\delta_{i,j+1} + \delta_{i, j-1}\right)$ is an $L \times L$ tridiagonal symmetric Toeplitz matrix.
Similarly, the fermionic representation of the charge $\Sigma$ is
\begin{equation}
	\Sigma =\sum_{j=L/2+1}^L (1/2 - a_j^\dagger a_j) =\sum_{i, j=1}^L a_i^\dagger \mathcal{S}_{ij} a_j,
\end{equation}
with ${\mathcal S}_{ij} = \langle i|\Sigma|j\rangle =  \frac{L}{4}\delta_{i,j} - \delta_{i,j > L/2}$.
The hermiticity of $H$ and $\Sigma$ is reflected in the symmetry of matrices $\mathcal{H}, \mathcal{S}$ of their fermionic representations.

\subsection{Trace formula}
\noindent
We now note a useful trace formula for exponentials of fermionic bilinears, similar to that derived by Klich \cite{klichJSM14}. The trace over the fermionic Fock space spanned by $|\underline n \rangle = |n_1, n_2, \ldots, n_L \rangle$ is given by
\begin{equation}
	{\rm Tr}[\ldots] = \sum_{\underline n} \langle \underline n | \ldots| \underline n \rangle.
\end{equation}
Start by considering the trace of the exponential of a fermionic bilinear generated by a matrix $A = \sum_{i,j=1}^L \mathcal{A}_{ij}a^\dagger_i a_j$.
\begin{equation}
	{\rm Tr}\left[e^A \right]={\rm Tr}\left[e^{\sum_{ij} \mathcal A_{ij}a_i^\dagger a_j} \right]. \label{ferm_trace}
\end{equation}
We rotate the matrix $\mathcal A$ to triangular form $\mathcal{R} \mathcal A \mathcal{R}^{-1} = \mathcal{D} + \mathcal{T}$ where $\mathcal{D} = {\rm diag}(d_1, \ldots, d_L)$ and $\mathcal{T}$ are diagonal and upper triangular matrices respectively. The fermionic trace in \eqref{ferm_trace} is now simple to evaluate
\begin{equation}
	{\rm Tr}\left[e^{\sum_{ij} \mathcal A_{ij}a_i^\dagger a_j} \right] = {\rm Tr}\left[e^{\sum_{i=1}^L \mathcal D_{ii}a_i^\dagger a_i} \right] = \prod_{j=1}^L (1+e^{d_j}) = \det \left[1+e^\mathcal{A}\right]. \label{ferm_trace_def}
\end{equation}
To extend the result to a product of exponential we note that
\begin{equation}
	[\mathcal A_{ij}^{(1)}a_i^\dagger a_j, \mathcal A^{(2)}_{kl} a^\dagger_{k}a_l] = [\mathcal A^{(1)}, \mathcal A^{(2)}]_{il}a^{\dagger}_ia_l,
\end{equation}
so that fermionic bilinears realize the representation of a Lie algebra. Now consider a product of exponentials of $A^{(1)}$ and $A^{(2)}$. Assuming that they can be represented as fermionic bilinears, the Baker–Campbell–Hausdorff formula gives
\begin{equation}
e^{\sum_{ij} \mathcal A^{(1)}_{ij}a_i^\dagger a_j}  e^{\sum_{ij}\mathcal A^{(2)}_{ij}a_i^\dagger a_j} = e^{\sum_{ij} \mathcal B_{ij}a_i^\dagger a_j},
\end{equation}
where $e^{\mathcal B} = e^{\mathcal A^{(1)}} e^{\mathcal A^{(2)}}$. Repeating the above argument and combining it with Eq.~\eqref{ferm_trace_def}, we come to the desired trace formula for products of fermionic bilinears
\begin{equation}
	{\rm Tr}\left[\overrightarrow \prod_p e^{\sum_{i,j=1}^L\mathcal A^{(p)}a_i^\dagger a_j } \right] = \det \left[1+ \overrightarrow \prod_{p} e^{\mathcal{A}^{(p)}} \right], \label{trace_formula}
\end{equation}
where $\overrightarrow{\prod}$ indicates the order of matrix multiplication. 

\subsection{Determinant form of the full counting statistics}
\noindent
We now consider the fermionic form of the brick-wall propagator $U$. Since the even and odd propagators \eqref{Ueo} factorize into commuting two-body maps, they can be written by splitting the continuous-time bilinear representation of the Hamiltonian \eqref{XX_ham_ferm} across alternating anti-diagonals as
\begin{equation}
	{U}_{\rm even} = e^{-\imath \delta t J \sum_{i,j=1}^L \mathcal{H}^{\rm e}_{ij}a_i^\dagger a_j}, \qquad
	{U}_{\rm odd} = e^{-\imath \delta t J \sum_{i,j=1}^L \mathcal{H}^{\rm o}_{ij}a_i^\dagger a_j}.
\end{equation}
Introducing even/odd Kronecker functions which distinguish parity of the first index
\begin{equation}
	\delta^{\rm e}_{ij} =
	 \begin{cases} \delta_{ij} & i\ {\rm even}\\
	 				0 & i\ {\rm odd}
	\end{cases}
, \qquad 	\delta^{\rm o}_{ij} = 
\begin{cases}
	0 & i\ {\rm even},\\
	\delta_{ij}  & i\ {\rm odd}
\end{cases}
\end{equation}
the splitting $\mathcal{H} = \mathcal{H}^{\rm e} + \mathcal{H}^{\rm o}$ reads
\begin{equation}
	\mathcal{H}_{ij}^{\rm e} =  \frac{1}{2}\left(\delta^{\rm e}_{i, j+1} + \delta^{\rm o}_{i, j-1} \right), \qquad
	\mathcal{H}_{ij}^{\rm o} = \frac{1}{2}\left(\delta^{\rm o}_{i, j+1} + \delta^{\rm e}_{i, j-1} \right).
\end{equation}
We now rewrite Eq.~\eqref{G_def} in terms of the fermionic trace, summing over repeated indices for clarity
\begin{equation}
	G(\lambda, T) =2^{-L}{\rm Tr}\left[\left( e^{-\imath \delta t J \mathcal{H}^{\rm e}_{ij}a_i^\dagger a_j}  e^{-\imath \delta t J \mathcal{H}^{\rm o}_{ij}a_i^\dagger a_j}\right)^T e^{\imath \lambda \mathcal{S}_{ij} a_i^\dagger a_j} \left( e^{\imath \delta t J \mathcal{H}^{\rm e}_{ij}a_i^\dagger a_j}  e^{\imath \delta t J  \mathcal{H}^{\rm o}_{ij}a_i^\dagger a_j}\right)^T  e^{-\imath \lambda  \mathcal{S}_{ij} a_i^\dagger a_j}\right].
\end{equation}
Applying the trace identity \eqref{trace_formula} we now find
\begin{equation}
	G(\lambda, T) = 
	\det M(\lambda, T) \label{LevitovLesovik_discrete_finiteL}
\end{equation}
with the matrix $M$ given by
\begin{equation}
	M(\lambda, T) = \frac{1}{2}\left(1 + \mathcal{U}^T e^{\imath \lambda \mathcal{S}} \mathcal{U}^{-T}  e^{-\imath \lambda \mathcal{S}}\right),
\end{equation}
where $\mathcal{U}=e^{-\imath \delta t J \mathcal{H}^{\rm e}} e^{-\imath \delta t J \mathcal{H}^{\rm o}}$ and $\mathcal{U}^{-T} = [\mathcal{U}^\dagger]^T$. Note that while Eq.~\eqref{G_def} involved tracing over an $2^L \times 2^L$ matrix, Eq.~\eqref{LevitovLesovik_discrete_finiteL} requires only the evaluation of a determinant of an $L\times L$ matrix. This reduction of dimension allows for efficient numerical evaluations of $G(\lambda, T)$ for much larger system and longer times.

\subsection{Cumulants}
\noindent
Finite-time cumulants $\kappa_n(T)$ are extracted as logarithmic derivatives of $G(\lambda, T)$ at the origin
\begin{equation}
	\kappa_{n}(T) = (-\imath)^n \partial_{\lambda}^n \log G(\lambda, T)|_{\lambda=0}. \label{c_def}
\end{equation}
Using Jacobi's formula $\partial_\lambda \det M= \det M\ {\rm Tr}[M^{-1} \partial_\lambda M]$ we can now derive explicit expressions for the first few lowest cumulants. We start by noting that
\begin{align}
	{M}^{-1}(0) = \mathds{1}, \qquad
	\partial^n_\lambda {M} = \frac{\imath^n}{2}\mathcal{U}^Te^{\imath \lambda \mathcal{S}}[\underbrace{\mathcal{S}, [\mathcal{S}, \ldots, [\mathcal{S}}_{n\ {\rm commutators}}, \mathcal{U}^{-T}]]]]e^{-\imath \lambda \mathcal{S}}.
	\label{M_der}
\end{align}
Evaluating at $\lambda =0$ and simplifying we have
\begin{equation}
	\partial^{n}_\lambda {M}|_{\lambda=0} = \frac{\imath^n}{2}  \mathcal{U}  \sum_{j=0}^{n}(-1)^j \binom{n}{j} \mathcal{S}^{n-j}\mathcal{U}^{-T}\mathcal{S}^{j}. \label{M_der_sum}
\end{equation}
It is straightforward to show that all odd cumulants identically vanish
\begin{equation}
	\kappa_{2n-1}(T) = 0.
\end{equation}
Introducing the notation $m_{p_1, p_2, \ldots} = {\rm Tr}\left[\partial_\lambda^{p_1} M(\lambda, T) \partial_\lambda^{p_2} M(\lambda, T) \ldots \right]_{\lambda=0}$ the lowest even cumulants are expressed as
\begin{align}
	\kappa_{2}(T) &= m_2 - m_{1,1}, \label{c2}\\
	\kappa_{4}(T) &= m_4 - 4m_{3,1}  - 3m_{2,2} + 12 m_{2,1,1} - 6m_{1,1,1,1}, \label{c4}\\
	\kappa_{6}(T) &= m_6- 6m_{5,1} - 15m_{4,2} + 30 m_{4,1,1} - 10m_{3,3} + 60m_{3,2,1} + 60m_{3,1,2} - 120m_{3,1,1,1} \\
	& \quad + 30m_{2,2,2} - 180m_{2,2,1,1} -90m_{2,1,2,1} + 360 m_{2,1,1,1,1} - 120m_{1,1,1,1,1,1}. \label{c6}
\end{align}
Note that while the above expressions for cumulants in terms of derivatives of the matrix $M$ are reminiscent of standard cumulants expressions in terms of moments, the expressions are distinct due to non-commutativity of the matrix terms. However Eq.~\eqref{M_der_sum} again expresses the derivatives in terms of traces of products of $L\times L$ matrices which allows for efficient numerical evaluations of the cumulants.

\newpage

\section{Asymptotic behavior of the correlation functions within BFT}

\noindent
Following Ref.~\cite{delvecchioJSM22}, the generating function for the distribution of the integrated current 
\begin{equation}
 \Omega = \int_0^t dt' j(0,t') 
\end{equation}
of a conserved quantity is defined as
\begin{equation}
 G(\lambda;t) = \left\langle e^{i\lambda\Omega(x,t)} \right \rangle.
\end{equation}
In the case of ballistic transport, in the limit 
$t \to \infty$, 
\begin{equation} 
 G(\lambda;t) \sim e^{Jt f_{\lambda}}
\end{equation}
up to possible power-law corrections, implying that 
\textit{all} cumulants scale linearly with time.
The BFT approach provides a way to calculate the rate function $f_\lambda$. For the XX spin chain and for the conserved $S^z$ (i.e., fermion number in terms of the Jordan--Wigner fermions) it is explicitly given by
\begin{equation} \label{eq:fasymp}
 f_{\lambda} = \frac14\int_{-\pi}^{\pi} \frac{dk}{2\pi} |v(k)| \log \left( \frac{1+e^{-w(\lambda;k)}}{1+e^{-w(k)}} \right),
\end{equation}
where the extra $1/4$ factor with respect to Ref.~\cite{delvecchioJSM22} accounts for the different conventions for the Hamiltonian (written in terms of the spin or the Pauli operators).
Here
\begin{equation}
 \begin{split}
  w(k) & = \beta \epsilon(k), \\
  w(\lambda;k) & = w(k) - \imath \lambda \,\sign[v(k)],
 \end{split}
\end{equation} 
where the dispersion relation and velocity are given by
\begin{equation}
 \begin{split}
  \epsilon(k) & = -4\cos(k), \\
  v(k) &= 4\sin(k).
 \end{split}
\end{equation}
For $T=\infty$ ($\beta=0$), the above expressions simplify
and \eqref{eq:fasymp} reduces to
\begin{equation}
f_{\lambda} = \int_{-\pi}^{\pi} \frac{dk}{2\pi} |v(k)| \log \left( \frac{1+e^{\imath \lambda \,\sign(k)}}{2} \right)
\end{equation}
which after carrying out the integration yields
\begin{equation}
f_{\lambda} = \frac{8}{\pi} \log\cos(\lambda/2).
\end{equation}


Given the form of the rate function, one realizes that \textit{all} odd cumulant vanish, and as mentioned above, \textit{all} even cumulants scale linearly with time.
In particular, this yields the following asymptotic scaling of the first few cumulants 
\begin{equation}
 \begin{split}
  \kappa_2 &\sim \frac{1}{2\pi} Jt, \\
  \kappa_4 &\sim -\frac{1}{\pi} Jt, \\
  \kappa_6 &\sim \frac{1}{2\pi} Jt, 
 \end{split}
\end{equation}
and for the corresponding standardized moments 
\begin{equation}
 \begin{split}
  \gamma_4 &= \kappa_4/\kappa_2^2 \sim -\pi (Jt)^{-1} \\ 
  \gamma_6 &= \kappa_6/\kappa_2^3 \sim 4\pi^2 (Jt)^{-2}.
 \end{split}
\end{equation}
Since all standardized moments beyond the variance are vanishing as $t\to\infty$, 
BFT predicts the distribution characterizing the FCS in the XX limit to be asymptotically Gaussian.

\newpage

\section{Additional numerical results}

\subsection{Numerical benchmark in the XX limit}
\noindent
In the XX limit ($\Delta=0$), the slow growth of the operator entanglement facilitates efficient simulations within the QGF framework. In contrast, QT exhibit poorer numerical performance at a significantly higher computational cost. 
In Fig.~\ref{fig:benchmark_XX} we show the quasi-continuous time evolution ($\delta tJ=0.1$) of the first few \textit{non-vanishing} cumulants and standardized moments. For $\kappa_2$ (a) and $\gamma_4 = \kappa_4/\kappa_2^2$ (b), 
the QGF reproduces the exact solution even for a low maximum bond dimension, $M = 2^6$. 
Calculating higher-order cumulants remains numerically challenging even in this simple limit 
due to the scaling $\kappa_k \propto |\lambda|^{-k}$. 
For example, evaluating $\gamma_6 = {\kappa_6/\kappa_2^3}$ for $\lambda = 0.03$ 
demands a precision (e.g., truncation error) of ${\cal O}(10^{-9})$. 
Increasing $\lambda$ could help but at the cost of a larger truncation error in the Taylor expansion of $G(\lambda, t)$ (not shown). 
The inset of Fig.~\ref{fig:benchmark_XX}(c) compares data for $\gamma_6$ with different bond dimensions. 
For $M=2^{6}$ the short-time oscillations of the exact solution are reproduced but not the long-time asymptotics.
A higher bond dimension significantly reduces the noise, as shown for $M=2^{10}$ in the inset of Fig.~\ref{fig:benchmark_XX}(c), 
but the simulation eventually fails on a (bond-dimension dependent) time scale $t_MJ$.

\begin{figure*}[hb]
\includegraphics[width=1.0\linewidth, angle=0]{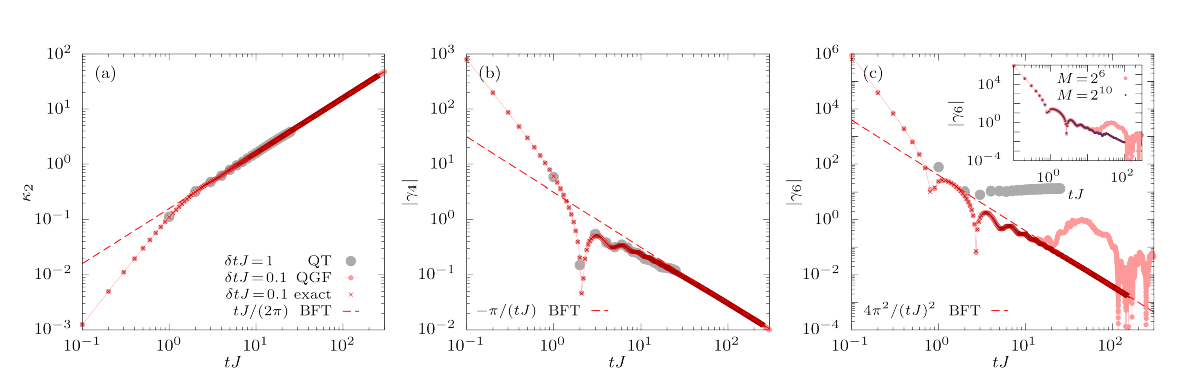}
\caption{Quasi-continuous time evolution of the variance $\kappa_2$ (a), 
kurtosis $\gamma_4=\kappa_4/\kappa_2^2$ (b), and 
sextosis $\gamma_6=\kappa_6/\kappa_2^3$ (c), in the XX limit ($\Delta=0$). 
The QGF results are validated against exact numerical results and asymptotic BFT predictions. 
Reproducing the correct beahvior of the sextosis at large times requires a higher precision, 
which can be achieved with a higher bond dimension, see inset of panel (c) or a larger time step, see Fig.~\ref{fig:floquet_XX}(c).
Parameters: $M=2^{6}$ (except for the inset), $\lambda=0.03$, $\delta tJ=0.1$ for QGF and $M=2^{10}$, $\delta tJ=1$ for QT. }
\label{fig:benchmark_XX}  
\end{figure*}

\begin{figure*}[hb]
\includegraphics[width=1.0\linewidth, angle=0]{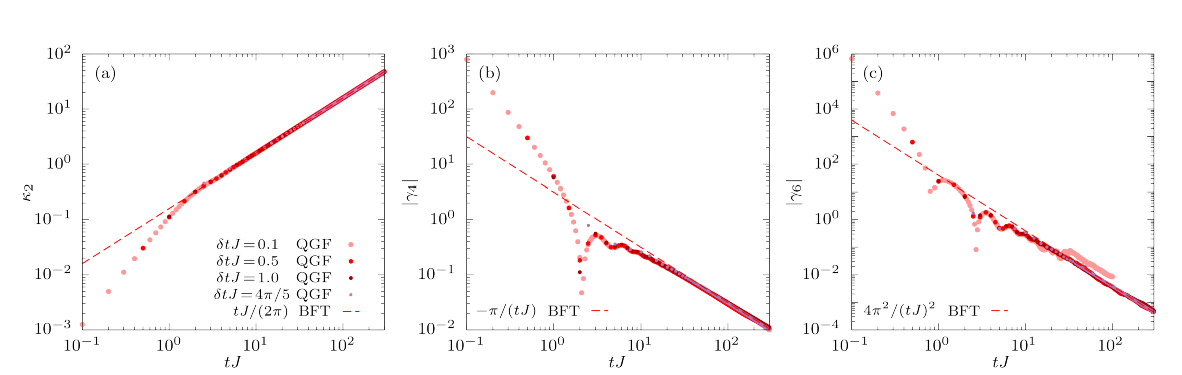}
\caption{Comparison of Floquet time evolution of the variance $\kappa_2$ (a), 
kurtosis $\gamma_4=\kappa_4/\kappa_2^2$ (b), and 
sextosis $\gamma_6=\kappa_6/\kappa_2^3$ (c), in the XX limit ($\Delta=0$). 
Increasing the time step $\delta t$ allows to achieve significantly better results for higher-order cumulants 
(specifically, for the sextosis) with respect to Fig.~\ref{fig:benchmark_XX}. 
Parameters: $M=2^{6}$ (for $\kappa_2$ and $\gamma_4$) and $M=2^{10}$ (for $\gamma_6$), $\lambda=0.03$, different $\delta tJ$ (as labeled). }
\label{fig:floquet_XX}  
\end{figure*}

\noindent
Significantly better results can be obtained by simulating faster dynamics. 
In Fig.~\ref{fig:floquet_XX} we show the Floquet time evolution of the same cumulant for increasing $\delta t$. 
Already at $\delta tJ=0.5$ the QGF data for $\kappa_6$ recover the exact asymptotic behavior. 
At $\delta tJ=1$ we find a minor accuracy improvement 
but a significant reduction of the computational resources required to reach $t_{\rm max}J=L/2$. 
At $\delta tJ=4\pi/5$ (as for the fSim gates in Google's experiment) the computational resources required are further reduced. 
Minor discrepancies in the short-time transient dynamics can be observed 
between Floquet and quasi-continuous time evolution, 
although for all values of $\delta tJ$ the simulations converge to the correct asymptotic behavior.

\FloatBarrier

\subsection{Floquet time evolution at the isotropic point}
\noindent
In constrat t the XX limit, in the superdiffusive regime, the kurtosis exhibits a notably different behavior 
between the quasi-continuous and Floquet time evolution dynamics, e.g., for the parameters used in the Google's experiment. 
This discrepancy calls for a more detailed investigation of the Floquet dynamics of the cumulants. 
In Fig.~\ref{fig:floquet_XXX} we show the time evolution 
of the variance $\kappa_2$ (a) and kurtosis $\gamma_4 = \kappa_4/\kappa_2^2$ (b) along the isotropic line ($\Delta=1$) 
for different time steps $\delta tJ$, as illustrated in the phase diagram (c). 
To minimize numerical noise in the kurtosis, we performed a subset of the simulations 
increasing the maximum bond dimension from $M=2^{10}$ to $M = 2^{12}$. 

\begin{figure*}[hb]
\includegraphics[width=1.0\linewidth, angle=0]{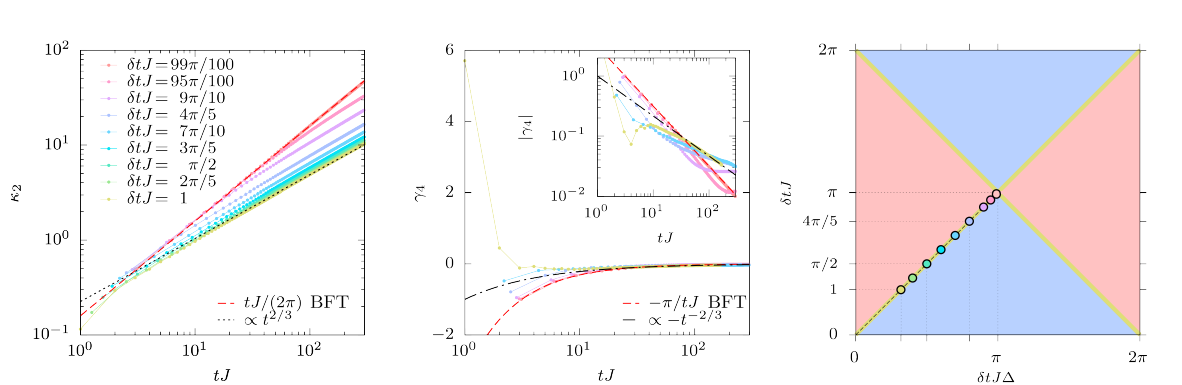}
\caption{Floquet time evolution of the variance $\kappa_2$ (a) and kurtosis $\gamma_4=\kappa_4/\kappa_2^2$ (b),  
along the isotropic line ($\Delta=1$) of the schematic phase diagram (c). 
There, the color-gradient symbols denote the corresponding time step $\delta tJ$ used for each simulation.  
The dashed red lines correspond to the BFT asymptotic prediction, which is recovered for trivial dynamics as $\delta tJ \to \pi$ (see text). 
The dashed black line $\kappa_2$ for denotes the superdiffusive scaling of the variance with dynamical exponent $z=3/2$, 
whereas the dot-dashed black line for $\gamma_4$ denotes an approximate empirical behavior in the quasi-continuous dynamics.   
Parameters: $M=2^{10}$ ($\kappa_2$) and $M=2^{12}$ ($\gamma_4$), $\lambda=0.03$ and different $\delta tJ$ (as labeled). }
\label{fig:floquet_XXX}  
\end{figure*}

\noindent
At the special point $\delta tJ \to \pi$ the local propagator reduces to a SWAP gate
\begin{equation}
 U_{j,j+1}=e^{\imath \delta t J h_{j,j+1} } \to
 \begin{pmatrix}
 1 & 0 & 0 & 0 \\
 0 & 0 & 1 & 0 \\
 0 & 1 & 0 & 0 \\
 0 & 0 & 0 & 1 
 \end{pmatrix} ,
\end{equation}
and the system displays a classical dynamics~\cite{ljubotinaPRL122a}, 
whereas $\delta t J=1$ is consistent with quasi-continuous time evolution. 
The QGF data for $\kappa_2$ displays a transient dynamics with a crossover between 
\textit{nearly}-ballistic ($z \approx 1$) and the asymptotic superdiffusive ($z=3/2$) regimes, 
with a characteristic time $t^{\star} \propto (\pi - \delta tJ)^{-\eta}$ for some exponent $\eta$. 
Already at $\delta tJ = 4\pi/5$ (which also corresponds to the time step of the fSim gates in Google's experiment) 
the crossover is barely noticeable. 
Note that there exists another transient regime where $\kappa_2 \sim t^2$ at times $tJ \lesssim 2$ 
which cannot be appreciated with such a fast Floquet dynamics 
but becomes apparent with a short-enough time step, see e.g., Fig.~\ref{fig:benchmark_XX}(a) 
but exists in \textit{all} regimes (i.e., for all investigated value of $\Delta$).

\noindent
A similar behavior is observed for the kurtosis $\gamma_4$. 
At $\delta tJ \approx \pi$ the kurtosis closely follows the BFT prediction for the XX limit to the longest time investigated,  
suggesting that it could be asymptotically vanishing as $\gamma_4 \sim -t^{-1}$. 
At shorter time steps, the kurtosis displays qualitative changes in its time evolution. 
The most striking features are: 
(i) a sign change at intermediate times (from positive to negative) observed for any time time steps $\delta tJ \lesssim 7\pi/10$,  
consistent with the data in Fig.2 (e) in the main paper and 
(ii) a seemingly different asymptotic behavior (see inset) which only converges towards 
(a speculative) $\gamma_4 \sim -t^{-2/3}$ for shorter time steps. 
However, despite reaching unprecedentedly long time scales, the data are not entirely conclusive, 
and we cannot exclude a universal asymptotic behavior to be recovered at sufficiently long time-scales.  

\FloatBarrier

\subsection{Scaling analysis of $G(\lambda,t)$ and $P(\Gamma,t)$}
\noindent
We turn to the scaling behavior of the generating function $G(\lambda,t)$ and of the full counting statistics $P(\Gamma)$ 
in the easy-plane limit (specifically, $\Delta=0$), the isotropic ($\Delta=1$), and easy-axis ($\Delta=2$) regimes.  

\noindent
In Fig.~\ref{fig:Glt_XXZ} we present time snapshots of $G(\lambda,t)$ obtained within the QGF and QT approaches.
In the XX limit (a), we find a nearly perfect agreement after the initial transient between QGF and QT data, 
whereas in the isotropic (b) and the easy-axis (c) regimes, the agreement deteriorates with time 
due to increasing truncation error. 
The QGF approach can reach significantly longer time-scales in all regimes. 
Each inset illustrates the universal collapse upon rescaling the counting field $\lambda \to \lambda t^{1/2z}$ 
with the appropriate dynamical exponent $z$ for each regime. 
The collapse is less effective in the easy-axis regime, where convergence to the dynamical exponent $z = 2$ is slow, as also indicated in the inset of Fig.~2(c) of the main paper.

\begin{figure*}[htp]
\includegraphics[width=1.0\linewidth, angle=0]{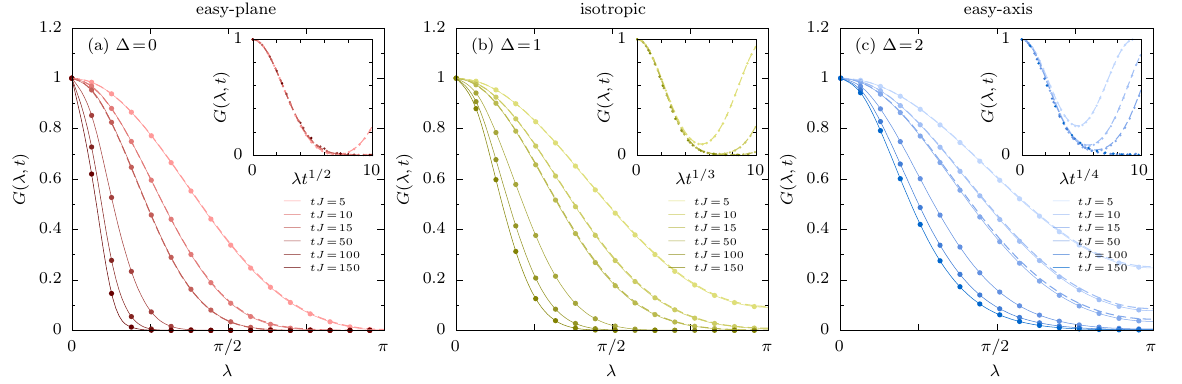}
\caption{Snapshots of the generating function at different times obtained within QT (dashed) and QGF (solid lines), 
and their scaling collapse (insets), 
obtained by rescaling the counting field $\lambda \to \lambda^{1/2z}$ with the appropriate dynamical exponent 
in the easy-plane (a), isotropic (b), and easy-axis (c) regimes. 
Parameters: $M=2^{6}$ (QGF at $\Delta=0$) and $M=2^{10}$ otherwise, $\delta t J =1$. }
\label{fig:Glt_XXZ}  
\end{figure*}

\begin{figure*}[htp]
\includegraphics[width=1.0\linewidth, angle=0]{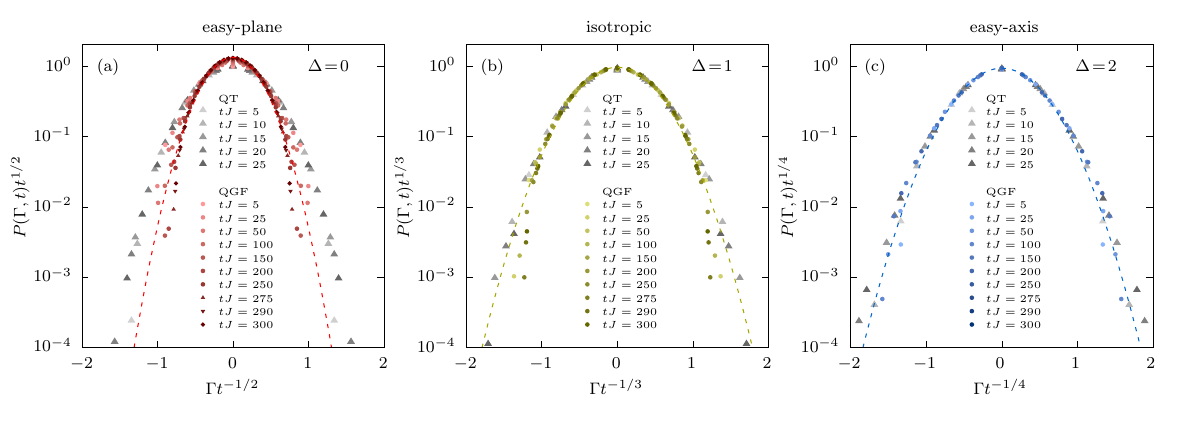}
\caption{Collapse of the full counting statistics with the dynamical exponent $t^{1/2z} P(\Gamma t^{-1/2z})$ 
for QT (grey gradient symbols) and QGF (color gradient symbols) in the easy-plane (a), isotropic (b), and easy-axis (c) regimes. 
Parameters: $M=2^{6}$ (QGF at $\Delta=0$) and $M=2^{10}$ otherwise, $\delta t J =1$. 
The dashed lines correspond to Gaussian distributions, for comparison.}
\label{fig:FCS_XXZ}  
\end{figure*}

\noindent
By performing the Fourier transformation of $G(\lambda,t)$ we can reconstruct $P(\Gamma,t)$. 
The resulting distributions are shown in Fig.~\ref{fig:FCS_XXZ} 
in the easy-plane ($\Delta=0$), the isotropic ($\Delta=1$), and easy-axis ($\Delta=2$) regimes. 
All distribution are distinctively symmetric, in agreement with the observation that 
all \textit{odd} cumulants identically vanish $\kappa_{2k+1}=0$. 
In each regime we observe a collapse of the curves when the distribution 
is properly scaled with the corresponding dynamical exponent as $P(\Gamma) \to t^{1/2z} P(\Gamma t^{-1/2z})$. 
In the XX limit (a), the scaled distribution asymptotically converges towards a Gaussian, 
which is expected to describe the asymptotic behavior for free fermions, 
and also in agreement with the observation that higher standardized moments asymptotically vanish $\gamma_{2n \ge 4} \to 0$. 
We ascribe the deviations observed along the tails, of the order $\mathcal{O}\left(10^{-3}\right)$,  
to numerical errors due to the Fourier transformation.  
In the other regimes, the QGF data show deviations from Gaussian distributions. 
t the longest time scale investigated, $P(\Gamma)$ collapses onto 
 a \textit{platykurtic} ($\gamma_4<0$) 
or \textit{leptokurtic} ($\gamma_4>0$) distribution 
in the isotropic (b) and easy-axis (c) regimes, respectively.  
The observed behavior of the FCS distribution corroborates the results obtained for the time evolution of the kurtosis,  
see Figs. 2(d,e,f) in the main text.


\end{document}